\documentclass[journal=mamobx,manuscript=article]{achemso}
\usepackage[version=3]{mhchem} 
\usepackage{graphicx}
\usepackage{caption}
\usepackage{subcaption}
\usepackage{natbib}
\usepackage{flexisym}

\author{Bekele Gurmessa}
\author{Robert Fitzpatrick}
\author{Tobias T. Falzone}
\author{Rae M. Robertson-Anderson}
\affiliation[USD]{Department of Physics and Biophysics, University of San Diego, San Diego, California 92110, United States}
\email {randerson@sandiego.edu}

\title {Entanglement Density Tunes Microscale Nonlinear Response of Entangled Actin}

\begin{document}

\begin{abstract}
We optically drive a microsphere at constant speed through entangled actin networks of 0.2 - 1.4 mg/ml at rates faster than the critical rate controlling the onset of a nonlinear response. By measuring the resistive force exerted on the microsphere during and following strain we reveal a critical concentration $c^{*}\simeq0.4$ mg/ml for nonlinear features to emerge. For $c>c^{*}$, entangled actin stiffens at short times with the degree of stiffening $S$ and corresponding timescale $t_{stiff}$ scaling with the entanglement tube density, i.e. $S\sim t_{stiff}\sim d_t^{-1}\sim c^{3/5}$. The network subsequently yields to a viscous regime with the yield distance $d_y$ scaling linearly with yield force $f_y$ and inversely with the entanglement length ($f_y\sim d_y\sim l_{e}^{-1} \sim c^{2/5}$). Stiffening and yielding dynamics are consistent with recent theoretical predictions for nonlinear cohesive breakdown of entanglements. We further show that above $c^{*}$ force relaxation proceeds via slow filament disengagement from dilated tubes coupled with $\sim$10x faster lateral hopping, with the corresponding concentration dependences in agreement with recent theoretical predictions for entangled rigid rods. 

\end{abstract}

\section{Introduction}
Actin, the most abundant protein in eukaryotic cells, is a semiflexible biopolymer in the cytoskeleton that plays a crucial structural and mechanical role in cell stability, motion and replication, as well as muscle contraction.  Most of these mechanically driven processes stem from the complex viscoelastic response that networks of entangled and crosslinked actin filaments display~\cite{stricker2010mechanics, wen2011polymer, gardel2008mechanical}. As such, numerous theoretical and experimental studies have been devoted to understanding the complex mechanics of this important biopolymer system~\cite{broedersz2014modeling, stricker2010mechanics}. Strained actin networks, characterized by a concentration-dependent mesh size of $\xi \sim 0.3c^{-1/2}$ (in $\mu$m, with concentration $c$ in mg/ml)~\cite{isambert1996dynamics}, exhibit wide-ranging mechanics which are highly dependent on  the concentration of filaments as well as the nature of the perturbation.  

Traditionally, polymer network mechanics have been studied via macrorheology techniques, which rely on bulk averaging and require often prohibitively large sample volumes for studies with valuable biopolymers such as actin. However, the emergence of microrheology techniques, which only require microliter sample volumes and directly probe molecular-level mechanics, has shed much needed new light onto the complex mechanics of actin networks at physiologically relevant force and length scales~\cite{stricker2010mechanics, janmey1994mechanical, kas1996f, koenderink2006high, hinsch2009non, semmrich2008nonlinear, liu2007visualizing}. Microrheology uses embedded microscale probes to sense network mechanics via tracking passively diffusing probes or using optical or magnetic tweezers to force the probe through the network while measuring the network response. While the nature of micro-and macro- rheology measurements are indeed quite different, several previous experimental and theoretical studies have been devoted to carefully determining the extent to which these two techniques can be compared~\cite{squires2005active, luan2008micro, buchanan2005comparing}. For example, we previously showed that for flexible polymers, active microrheology reported macroscopic properties for probe sizes $>$ 3x the tube diameter of the network~\cite{chapman2014onset, mcleish2002tube, cai2011mobility}. Nonetheless, flows induced by  microrheology are in fact non-uniform, and can lead to polymer buildup at the leading edge and depletion at the trailing edge of a moving probe leading to unique microscale visoelastic regimes not present at the macroscale~\cite{falzone2015entangled, uhde2005osmotic}, leading to unique microscale viscoelastic regimes not present at the macroscale\cite{uhde2005osmotic,uhde2005viscoelasticity}. However, it is these discrepancies between the two techniques that can actually shed light on the microscale structures of biopolymer networks and varying lengthscale-dependent mechanical regimes~\cite{falzone2015entangled, squires2008nonlinear}. Despite these technological advances, most of the previous rheological studies on actin networks have focused on the near-equilibrium linear regime (accessible to passive microrheology techniques) and mechanics of cross-linked networks~\cite{mason1995optical, gardel2003microrheology, liu2006microrheology}. As such, the microscale response of entangled actin subject to nonlinear strains which perturb the network far from equilibrium, has remained largely unexplored~\cite{falzone2015entangled, falzone2015active}.   

The dynamic properties of entangled polymer solutions are tuned primarily by their entanglement density or concentration, and have successfully been described by the well celebrated tube model pioneered by de Gennes and Doi and Edwards~\cite{de1979scaling, doi1988theory}. This model, which postulates that each polymer chain is constrained to a tube-like region formed by the surrounding chains that prohibits motion transverse to its contour, was originally developed for flexible polymer chains and rigid rods. Actin with a persistence length, $l_p$, of ~17 $\mu$m, comparable to typical filament lengths of $\sim$10 - 30 $\mu$m, falls in a unique class of semiflexible polymers that are able to bend and deflect but maintain extended profiles~\cite{liu2006microrheology, odijk1983statistics,kas1994direct}. Historically, researchers have often applied rigid rod reptation theory to entangled actin networks with reasonable success~\cite{janmey1994mechanical,kas1996f}. For example, the scaling for the reptation time ($\tau_{D}$) and the diffusion constant along the tube ($D_{||}$) with filament length, have been shown to agree with classic Doi-Edwards (DE) theory for rigid rods~\cite{kas1994direct, kas1996f}. According to rigid rod DE theory the longest relaxation time of the network is the reptation time, the time it takes for a filament to completely diffuse out of its tube, which depends both on the tube diameter, $d_t$, and filament length $L$ as $\tau_{D} \sim $ ($L/d_t)^2$ \cite{doi1988theory}. To account for discrepancies, such as $D_{||}$ values $\sim$20 smaller than predicted for rigid rods~\cite{kas1996f}, wormlike chain models were developed that incorporated bending and lateral deflection modes into tube models and placed more significance on the persistence length as a controlling lengthscale for network dynamics~\cite{tracy1992dynamics, odijk1986theory, kas1996f, isambert1996dynamics, maggs1998micro}. Within this framework the governing network lengthscales are the tube diameter $d_t \equiv \xi^{6/5}l_p^{-1/5}$ $\sim c^{-3/5}$, the filament length between entanglements $l_e \equiv \xi^{4/5}l_p^{1/5} \sim c^{2/5}$, the Odijk deflection length $\lambda \equiv \xi^{2/3}l_p^{1/3} \sim c^{-1/3}$  and the amplitude of transverse fluctuations $l_t^2 \sim L^3/l_p$~\cite{isambert1996dynamics, odijk1983statistics, maggs1990adsorption}. Given these lengthscales, deformed actin filaments are predicted to release their stress via several dynamically distinct mechanisms that occur over different length and time scales dependent on the network concentration~\cite{isambert1996dynamics, gardel2003microrheology, hinner1998entanglement}. 

Initial relaxation occurs when a mesh size length along a filament can relax. This mesh time, $\tau_{mesh} \equiv \beta\zeta\xi^4l_p^{-1}$, where $\beta = 1/k_{B}T$ and $\zeta$ is the translational friction coefficient~\cite{isambert1996dynamics, odijk1986theory}, is the time it takes for hydrodynamic interactions between filaments to become important (i.e. for each filament to ``feel" the network). The second fastest relaxation mechanism is the relaxation of individual entanglement segments which occur over the disentanglement time $\tau_{ent} \equiv \beta\zeta\xi^{16/5} l_p^{-1/5}$, and is only weakly dependent on the rigidity of the polymer.~\cite{isambert1996dynamics} Odijk length fluctuations, on the other hand, unique to semiflexible filaments, relax on a timescale $\tau_{\lambda} \sim \beta\zeta L \lambda^2 \sim c^{-2/3}$. Complete terminal relaxation is only predicted to occur after the confined polymer can completely diffuse or disengage from the deformed tube. This predicted reptation time $\tau_{D} \sim \beta\zeta L^3$ is concentration independent (ignoring any concentration dependence of $\zeta$), in contrast to the rigid rod result $\tau_{D} \sim (L/d_t)^2 \sim c^{6/5}$. Determining exact numerical quantities for these timescales requires knowing the friction coefficient $\zeta$. While $\zeta$ is often assumed to be concentration independent (for rigid rods) with $\beta\zeta = 1/(D_{||}L) \sim c^0$, effective medium arguments have shown a weak increase in $\zeta$ with concentration (i.e. $\zeta \sim 1/(ln (c^{-1/2})))$~\cite{muthukumar1983screening, odijk1986theory} and with a power-law fit to the data presented by Kas et al. for diffusion of entangled actin filaments~\cite{kas1996f} one can demonstrate a similar weak increase, $\zeta \sim c^{~0.35}$. If one assumes no concentration independence, estimated values of $\beta\zeta \simeq 2-4  s/{\mu}m^3$ can be determined from previously reported data and predictions~\cite{isambert1996dynamics, kas1996f}.   

Notably, these relaxation timescales are all for steady-state or linear regime mechanics and the extent to which bending, retraction and semi-flexibility play a role in nonlinear mechanics of entangled actin remains controversial. A recent theoretical extension to the tube model by Sussman and Schweizer (SS), describing the nonlinear response of entangled rigid rods, shows agreement with several previous entangled actin experiments~\cite{wang2010confining, glaser2010tube, robertson2007direct}, suggesting that bending, and stretching play a negligible role in the nonlinear regime. Further, our recent microrheology measurements show that nonaffine filament displacements and stress-softening due to bending modes are suppressed in the nonlinear regime~\cite{falzone2015active}. In fact, classical DE theory predictions for nonlinear mechanics are largely the same for flexible and rigid polymers due to the fast equilibration of chain stretch via Rouse modes~\cite{doi1988theory, sussman2012microscopic, sussman2013entangled}.  

We finally note that tube models and extensions there of have all been developed for monodisperse entangled polymer systems, whereas networks of actin as well as most synthetic polymers have a distribution of lengths around the mean. However, these models have routinely been applied to both systems, showing excellent agreement between theory and experiment for most steady-state and linear regime properties~\cite{kas1994direct, wang2010confining}, suggesting that the degree of polydispersity in these systems does not play a significant role in mechanics.
Using the tube model framework several theoretical studies~\cite{frey1998viscoelasticity, maggs1998micro, morse1998viscoelasticity} have predicted that the plateau modulus scales with concentration as $G \sim c^{7/5}$ while others~\cite{morse2001tube} predict $G \sim c^{4/3}$. Several macro- and microrheology studies have reported agreement with the $c^{7/5}$ scaling~\cite{gardel2003microrheology, hinner1998entanglement, palmer1999diffusing}, and recent microrheology measurements of the microscale creep compliance of entangled actin reported that viscosity scales as $\eta\sim c^{7/5}$~\cite{uhde2005viscoelasticity}.  On the other hand, other recent microrheology measurements~\cite{koenderink2006high, xu1998rheology, palmer1999diffusing} reported scaling of $G\sim c$ while still others~\cite{janmey1991viscoelastic, mackintosh1995elasticity, gardel2004elastic, atakhorrami2014scale} showed $G\sim c^{11/5}$ and $G\sim c^{9/16}$. Further, the majority of experimental and theoretical studies have reported stress softening of entangled actin rather than stiffening in the nonlinear regime, understood to be due to the available non-affine bending modes not present in crosslinked or rigid rod networks~\cite{stricker2010mechanics}. However, two previous macrorheology studies have reported stress-stiffening at short times for entangled actin~\cite{semmrich2008nonlinear, xu2000strain}. These discrepancies in both linear and nonlinear regimes demonstrate the lack of clarity regarding the concentration dependence of entangled actin mechanics, and highlight the need for new experiments that systematically characterize the concentration-dependent response of entangled actin especially in the nonlinear regime. 

The classic Doi-Edwards assumption that the entanglement tube is a static intrinsic network parameter~\cite{doi1988theory}, has been remarkable in predicting linear regime mechanics of entangled polymers, yet has failed to accurately describe nonlinear response features such as stress-stiffening, shear thinning, and entanglement tube dilation~\cite{semmrich2008nonlinear, xu2000strain, sussman2012microscopic, sussman2011communication}. This failure lies in its poor treatment of chain disentanglement during large mechanical deformations that lead to nonlinear responses~\cite{wang2013new}. The mounting experimental evidence of complex entanglement dynamics and stress nonlinearities not captured by the DE model~\cite{sussman2012microscopic, sussman2011communication}, has led to the emergence of tube model extensions such as convective constraint release (CCR), cohesive yielding, stress-induced tube dilation, and anharmonic tube potential softening, all of which dynamically alter the density and propensity of entanglements surrounding each filament~\cite{mhetar1999nonlinear, sussman2012microscopic, sussman2011communication, lu2014origin, wang2007new}. CCR models postulate that the imposed shear forces surrounding entanglements to disentangle at rate controlled by the shear rate, allowing each filament to relax over a much faster timescale than reptation~\cite{mcleish2002tube}. Recent theoretical considerations of Wang et al.~\cite{wang2007new, wang2012diffusion} suggest that stress stiffening and subsequent yielding arises as the induced stress becomes greater than the cohesive entanglement force. This cohesive force $f_{coh}$, resulting from the entanglements along each polymer restricting relaxation, is predicted to scale inversely with $l_{e}$ (or proportional to the entanglement density along each polymer). The force or stress that induces yielding or cohesive breakdown, $f_{y}$, which is comparable to $f_{coh}$ at the yield point, scales linearly with the yield strain or distance (i.e. $f_{y} \sim d_{y} \sim l_{e}^{-1}$. The recent SS tube model extension ~\cite{sussman2012microscopic, sussman2013entangled, sussman2011communication} which assumes that the topological constraint that the entangled chains impose (i.e. the tube) is directly dependent on the externally applied stress, leads to strain-induced tube dilation arising from alignment of filaments with the shear which reduces the probability of inter-filament collisions (i.e. entanglements), thereby widening the tube (or reducing the entanglement density). Models of SS and Wang~\cite{wang2007new} also predict tube potential softening and yielding, in line with previous entangled actin~\cite{wang2010confining} and DNA measurements~\cite{robertson2007direct}, in which the potential becomes weaker over time and transitions from harmonic to quasilinear as fluctuations approach $d_{t}$. This stress-induced dilation and yielding, which ultimately leads to faster network relaxation, is predicted to be concentration dependent. Specifically, tube dilation from its equilibrium size $d_t$ scales as $d_t\textprime/d_t \sim c^{1/2}$ where $d_t\textprime$ is the dilated tube diameter~\cite{sussman2012microscopic,sussman2013entangled}. Provided the relation $\tau_{D} \sim d_t^{-2}$, the decrease in $\tau_D$ ($\tau_D\textprime/\tau_D$) is then predicted to scale inversely with concentration (i.e. $\tau_D\textprime/\tau_D \sim c^{-1}$). Further, a second relaxation mechanism, transverse barrier hopping, is predicted, whereby filaments can evade tube confinement as fluctuations lead to temporary removal of the potential. The rate of this alternative relaxation mechanism is likewise inversely dependent on concentration, and becomes competitive with and eventually faster than the reptation time for strains larger than the yield strain. Thus, tube dilation and yielding can collectively lead to a two phase relaxation following strain which is faster than single mode relaxation of filaments out of un-dilated tubes (classic DE relaxation). Despite such theoretical progress on nonlinear mechanics, the lack of corroborating experiments coupled with a dearth of nonlinear theories specifically designed for semiflexible filaments, leaves the nonlinear mechanics of entangled actin networks still not well understood. Further, very few of these theoretical extensions have incorporated concrete predictions regarding the concentration dependence of mechanics. 

We previously used optical tweezers microrheology to characterize the microscale force response of entangled actin of a single concentration (0.5 mg/ml) subject to constant strain deformations~\cite{falzone2015entangled}. We demonstrated a unique strain rate dependent crossover from linear to nonlinear mechanics that occurred at a strain rate of $\dot\gamma_c \sim 3 s^{-1}$. For strains faster than $\dot\gamma_c$, actin networks exhibited stress-stiffening, shear thinning, and strain induced entanglement tube dilation. Tube dilation dynamics, manifested during the post-strain force decay as two-phase force relaxation, were consistent with the SS tube model extension for rigid rods. These unique observations for a fixed actin concentration motivated the question as to how robust this nonlinear regime is. Namely, how entangled must the filaments be for this nonlinear response to emerge? How do dynamics scale with concentration in the nonlinear regime? 

Here we use active optical tweezers microrheology to directly probe the microscale mechanical properties of entangled actin networks over a broad range of concentrations (0.2 - 1.4 mg/ml). We drive the probe at rates faster than $\dot\gamma_c$, to nominally access the nonlinear regime. We show that the onset of nonlinearity is not simply controlled by strain rates but is intrinsically linked to the entanglement density (concentration) of the network. Only for networks in which the length between entanglements is $\leq1 \mu$m, corresponding to a concentration of $c^{*}\simeq 0.4 $ mg/ml, are filaments restricted enough to display nonlinear mechanics. For concentrations above $c^{*}$, we observe a break from predicted linear scaling laws coupled with key nonlinear features including concentration-dependent strain-stiffening, yielding, shear thinning, and entanglement tube dilation. Our measurements are the first to comprehensively characterize the concentration dependence of the nonlinear force response of entangled actin filaments at the microscale. Our results, directly applicable to the larger class of semiflexible polymers, corroborate recent theoretical predictions as well as reveal dynamical scaling laws not previously identified. Thus, these measurements fill a long-standing gap in knowledge regarding the microscale dynamics of entangled polymers driven far from equilibrium.   

\section{Materials and Methods}

Unlabeled and Alexa-568-labeled rabbit skeletal muscle actin were purchased from Cytoskeleton (AKL99) and Invitrogen (A12374), respectively. Actin was stored in Ca Buffer G [2mM Tris pH 8.0, 0.2mM ATP, 0.5mM DTT, 0.1mM $\mathrm{CaCl_2]}$ and polymerized for 1 hour in F-buffer [10mM Imidazole pH 7.0, 50mM KCl, 1mM $\mathrm{MgCl_2}$, 1mM EGTA, 0.2mM ATP]. 4.5 $\mu$m carboxylated  polystyrene microspheres (probes, Polysciences Inc.) were labeled with Alexa-488 BSA (Invitrogen) to inhibit interaction with the actin network\cite{valentine2004colloid} and visualize the probes during measurement. Actin networks for experiments were generated by mixing labeled actin, unlabeled actin, and probes, in F-buffer for final actin concentrations of $c$ = 0.2  - 1.4 mg/ml. These concentrations correspond to $\xi = 0.67 - 0.25$ $\mu$m, $d_t = 0.35 - 0.11$ $\mu$m, and $l_e = 1.28 - 0.59$ $\mu$m. The mixture was quickly pipetted into a sample chamber made from a glass slide and cover slip separated $\sim$100 $\mu$m by double sided tape such that it can accommodate $\sim$ 30 $\mu$L solution and sealed with epoxy. 
Actin filament length can vary at different polymerization concentrations, so to quantify filament lengths and incorporate potential length variations into our interpretations, we imaged each network using a Nikon A1R confocal microscope prior to each experiment ~\cite{falzone2015entangled, kasza2010actin}. For each concentration $\sim$500 filaments were measured to obtain length distributions (SI Fig 1). The length distributions displayed minimal dependence on concentration with lengths of 7.8 $\pm$ 3.3 and 7.9 $\pm$ 2.4 for 0.2 and 1.4 mg/ml respectively, resulting in an average filament length 7$\mu$m. To ensure that the slight concentration-dependent length variation did not play a role in our measurements, we also repeated measurements using the capping protein gelsolin (Cytoskeleton,inc. (HPG6)) to fix actin length to 7 $\mu$m (using the relation L = $(330R_G)^{-1}$ where $R_G$ is the molar ratio of gelsolin to actin)~\cite{kasza2010actin}. There was minimal difference between measured lengths and force measurements for filaments polymerized with and without gelsolin (SI Fig 2).

We note that the filament lengths used in this study are $>$2x smaller than the persistence length and $\sim$2-4x smaller than typical filament lengths used in previous studies ($\sim$20-40 $\mu$m) that have shown discrepancies between rigid rod theories and experimentally measured actin mechanics~\cite{kas1996f, koenderink2006high, isambert1996dynamics}.

\begin{figure}[ht!]
\centering
\includegraphics[height=0.44\textwidth]{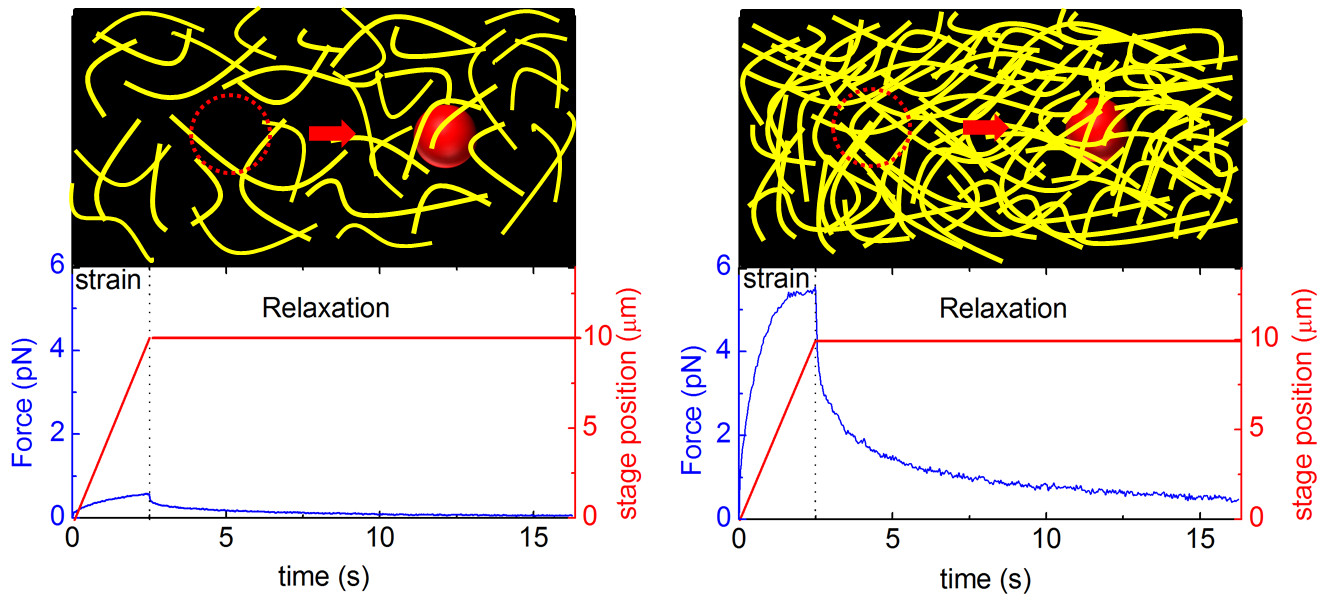}
\caption{Typical experimental setup. Top panels depict a 4.5 $\mu$m optically trapped probe (red) displaced $10 ~\mu$m at constant speed (4 $\mu$m/s and 8 $\mu$m/s) through an entangled actin network of low (left) and high (right) concentration. The bottom panels display the corresponding measured resistive force the actin exerts on the probe (blue) both during (strain) and following (relaxation) the probe motion (shown in red). The specific force and position data shown are for 0.2 mg/ml (left) and 1.0 mg/ml (right) actin subject to a 4 $\mu$m/s strain.}\label{Actin0}
\end{figure}

 The optical trap used in measurements was formed by a 1064 nm Nd:YAG fiber laser (Manlight) focused with a 60x 1.4 NA objective (Olympus). A position-sensing detector (Pacific Silicon Sensors) measured the deflection of the trapping laser, which is proportional to the force acting on the trapped probe over our entire force range. The trap stiffness was calibrated via Stokes drag in water~\cite{williams2002optical} and passive equipartition methods~\cite{brau2007passive}.  During measurements, a probe embedded in the network is trapped and moved 10 $\mu$m at a constant speed (4 $\mu$m/s and 8 $\mu$m/s) relative to the sample chamber via translation of a nanopositioning piezoelectric stage (Mad City Labs) while measuring both the laser deflection and stage position at a rate of 20 kHz during the three phases of experiment: equilibration ($15 ~s$), strain ($1.25 ~s$ and $2.5 ~s$) and relaxation ($15 ~s$) (Fig.~\ref{Actin0}). By using the established conversion factor for microrheology measurements in complex fluids $\dot\gamma = \frac{3\nu}{\sqrt{2}r}$, where \textit{v} and \textit{r} refer to the speed and radius of the probe~\cite{falzone2015entangled, squires2008nonlinear}, our chosen speeds correspond to strain rates of $\dot\gamma$ = $3.8~s^{-1}$ and $7.5~s^{-1}$. These rates were chosen to be comparable to and faster than our previously determined crossover rate of $\dot\gamma_c \simeq 3 s^{-1}$ that controls the onset of nonlinear mechanics. Precision measurements and data acquisition were achieved using LabVIEW while custom-written MATLAB programs were used for post-measurement data analysis. Displayed force curves (Figs 2-4) are averages of $50$ trials using $50$ different probes each at different locations in the sample chamber. All error bars were determined by bootstrapping over $1000$ subsets.

\section{\textbf{Results and discussion}}

\begin{figure}[ht!]
\centering
\includegraphics[height=0.8\textwidth]{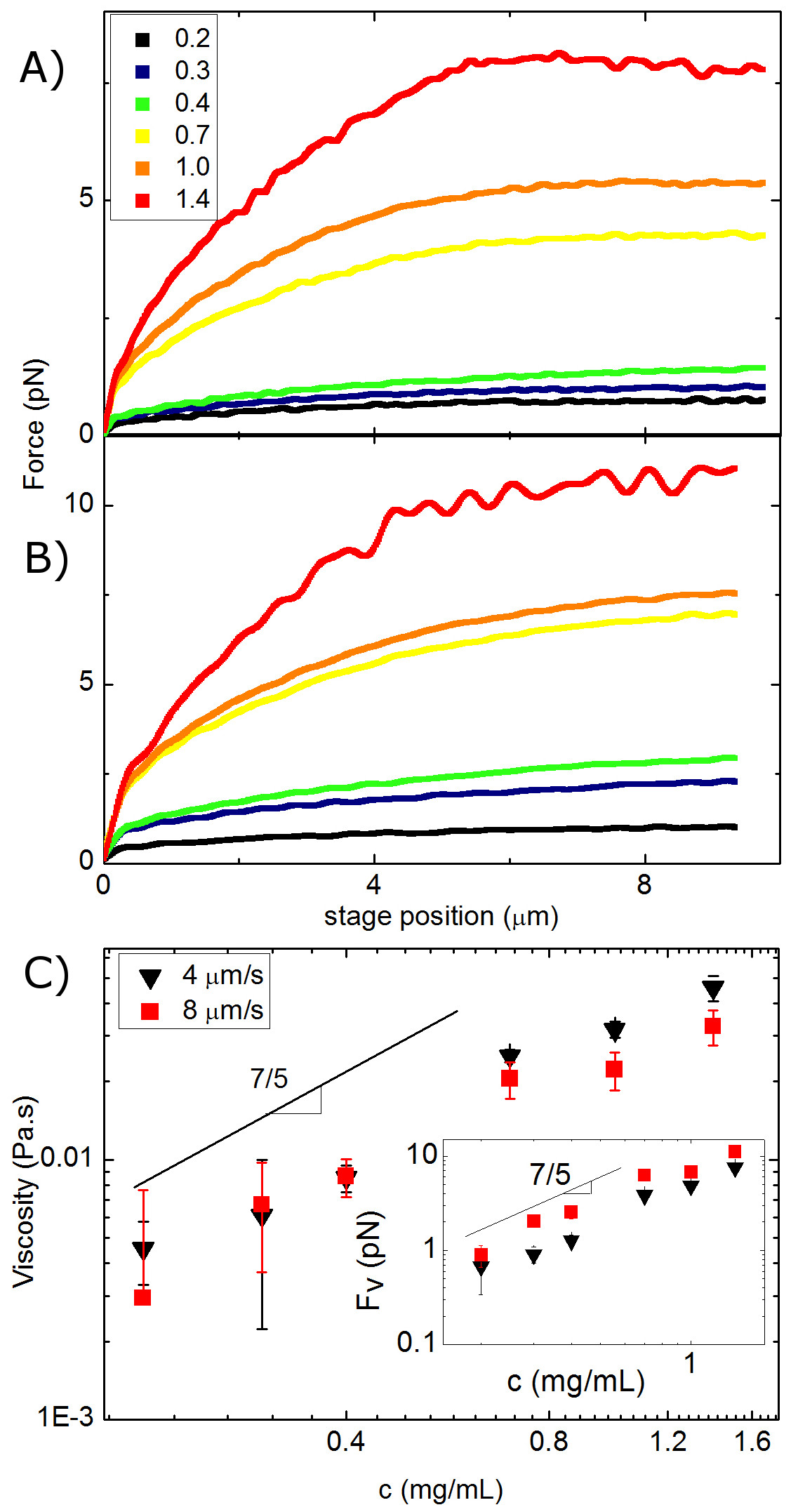}
\caption{Microscale force response of entangled actin displays shear thinning and breakdown of linear regime mechanics for concentrations above 0.4 mg/ml. (A,B) Average force exerted by entangled actin networks of varying concentrations (mg/ml, listed in legend) to resist a probe driven at constant speeds of (A) 4~$\mu$m/s and  (B)  8~$\mu$m/s. (C) Effective viscosity as a function of concentration determined from the corresponding plateau force $F_v$ reached at the end of the strain (shown in inset). The solid line shows the previously reported scaling $\eta\sim c^{7/5}$ for linear regime mechanics of entangled actin. Note that at higher concentrations the viscosity deviates from this scaling and decreases with increasing speed suggesting the onset of shear thinning. The inset shows the resistive force from which the viscosity is determined and demonstrates agreement with the scaling predicted for the plateau modulus ($G\sim c^{7/5}$) at low concentrations ($c < c^{*}$). However, this agreement breaks down for $c>c^{*}$, indicating tube dilation and shear thinning}\label{Actin1}
\end{figure}


Figure~\ref{Actin1} (A,B) illustrates the measured force entangled actin filaments of varying concentrations exert on the probe to resist the $10 ~\mu$m strain. For all concentrations and speeds actin networks initially respond elastically, with resistive force increasing with strain distance, after which the force approaches a strain-independent plateau indicative of a purely viscous response. Using this strain-independent plateau force, $F_v$, and Stokes' drag, $\eta = F_v/6\pi r v$,  (Fig.~\ref{Actin1}C), we show that for the three lower concentrations the effective viscosity is speed-independent, representative of a Newtonian viscous response in which the viscosity is an intrinsic property of the fluid unaffected by applied strain. In this low concentration regime the viscosity scales as $\eta \sim c^{7/5}$ in agreement with previous microrheology creep response measurements of entangled actin at concentrations of $0.1 - 1.4$ mg/ml~\cite{uhde2005viscoelasticity}. This unique micro-viscosity scaling is suggested to arise from a steady-state build up of filaments in front of the probe and depletion behind the probe, rather than shear friction dissipation as in macrorheology measurements.~\cite{uhde2005viscoelasticity}
In contrast, for the higher concentrations, the viscosity not only deviates from the 7/5 scaling, but also decreases with increasing speed, indicating the onset of shear-thinning. Namely, viscosity is no longer an intrinsic property of the network, rather the network viscosity is rate-dependent, becoming less viscous with increasing speed. As described above, shear-thinning is a nonlinear response of well-entangled polymer networks subject to large strains, and can be understood as a result of strain-induced disentanglement (CCR) or filament alignment with the shear (leading to a reduced entanglement density or tube dilation). We previously reported similar shear-thinning (specifically $\eta \sim \dot\gamma^{-1/3}$) for $0.5$ mg/ml actin subject to nonlinear strains ($\dot\gamma > \dot\gamma_c$), indicating that concentrations below 0.5 mg/ml are not sufficiently entangled to exhibit nonlinearities. As shown in the inset of Fig.~\ref{Actin1}C, we also examine the concentration dependence of the steady-state force response $F_v$. In the linear regime, the steady-state force should scale with the linear shear modulus of the network $G$ and the strain speed $v$. For $c<c^{*}$ we find $F_v$(8~$\mu$m/s) $\sim $ 2$F_v$ (4 ~$\mu$m/s) with a scaling with concentration in agreement with the previous theoretically predicted shear modulus scaling of $c^{7/5}$~\cite{hinner1998entanglement, frey1998viscoelasticity}. Tube dilation would lower this scaling exponent, as the effective shear-induced concentration (entanglement density) is lower than the nominal value; and shear thinning would reduce the linear dependence of $F_v$ on speed, as the viscosity would be reduced in a rate-dependent manner. Indeed for $c>c^*$ we find that $F_v$ is nearly concentration and speed independent for concentrations above 0.4 mg/ml.

We also quantify the time at which the network yields to the viscous-dominated response by determining the time at which the differential modulus, or the derivative of force with position ($K = dF/dx$), becomes negligible ($\frac{1}{2e}$ of initial $K$; see below). As shown in Fig.~\ref{Actin2}D, for $c>c^{*}$ we find that the distance or strain at which yielding occurs $d_{y}$ scales with concentration as $d_{y} \sim c^{2/5}$. As $l_{e} \sim c^{-2/5}$, this scaling implies that the yield distance is proportional to the number of entanglements along each filament. This scaling is consistent with the cohesive breakdown model of yielding developed by Wang et al, in which $f_{y} \sim l_{e}^{-1}$. Within this model, $f_{y} \sim d_{y}$ just as we see in the inset of Fig ~\ref{Actin2}D. As described in the introduction, this yielding arises when the induced force balances the cohesive elastic force provided by the entanglements restricting filament motion~\cite{wang2007new,wang2012diffusion}.

To further quantify the initial elastic force response of entangled actin, we calculate an effective differential modulus ($K=dF/dx$) which quantifies the elasticity or stiffness of the network. In other words, an increase in $K$ signifies stress-stiffening while a decrease signifies stress-softening and ultimately yielding to a purely viscous response ($K \simeq 0$). As displayed in Figure~\ref{Actin2}(A,B), we find that entangled actin initially stiffens, increasing from an initial value ($K_0$) to a maximum ($K_{max}$), and subsequently softens to a viscous regime. While nearly all concentrations exhibit some degree of stiffening ($S = (K_{max}/K_0) - 1$), albeit minimal for $c < 0.4$ mg/ml, stiffening increases with concentration and is independent of speed (Fig.~\ref{Actin2}(C)). If stiffening is a result of entanglement segments resisting deformation, then we should expect $S$ to scale with the number of entanglements $N_e$ that the probe passes through during strain ($S \sim N_e$). Because the probe always moves a fixed distance ($10~\mu$m) then the number of entanglements that it encounters will increase with concentration. Specifically, if $d_t \sim c^{-3/5}$ then $N_e\sim c^{3/5}$ which is nearly the scaling of $S$ that we find for higher concentrations. The speed-independence of stiffening demonstrates that stiffening is indeed controlled by intrinsic length scales of the network (i.e. $d_t$). This scaling further supports the concept of a cohesive force on each filament, which is proportional to the number of entanglements constraining its motion, that provides the elastic response. The timescale over which stiffening occurs also scales with $N_e$ which can be understood as a stronger cohesive force enabling elastic stiffening to persist for longer times. In other words, the higher $N_e$ is the larger the strain must be to rupture entanglements and suppress further stiffening.  Further, while $t_{y}$ scales with $N_e$, the timescales for all concentrations are comparable to the theoretically predicted $\tau_{mesh}$ ($\sim$0.01 s) as seen in the inset of Fig.~\ref{Actin2}(C), and is in line with our previous findings for 0.5 mg/ml actin. Because the fastest relaxation timescale of the network is the mesh relaxation time, for $t < \tau_{mesh}$ the network cannot undergo any relaxation or softening, and thus stiffens in response to the strain. Specifically, such stiffening has previously been suggested to arise from alignment of entanglement segments in the direction of strain, suppressing non-affine bending modes, leading to a costly stretching-dominated response~\cite{onck2005alternative, semmrich2008nonlinear, xu2000strain, xu1998rheology}. After a time $\tau_{mesh}$, each filament is hydrodynamically constrained by the surrounding filaments, prohibiting further free rotation to orient with the strain, thereby suppressing further stiffening. Once the network can begin to relax (via filament interactions) the force response can become more viscous (i.e. the network can flow) so the force response exhibits softening and yielding (Fig~\ref{Actin2}(A,B)). 

\begin{figure}[ht!]
\centering
\includegraphics[height=0.5\textwidth]{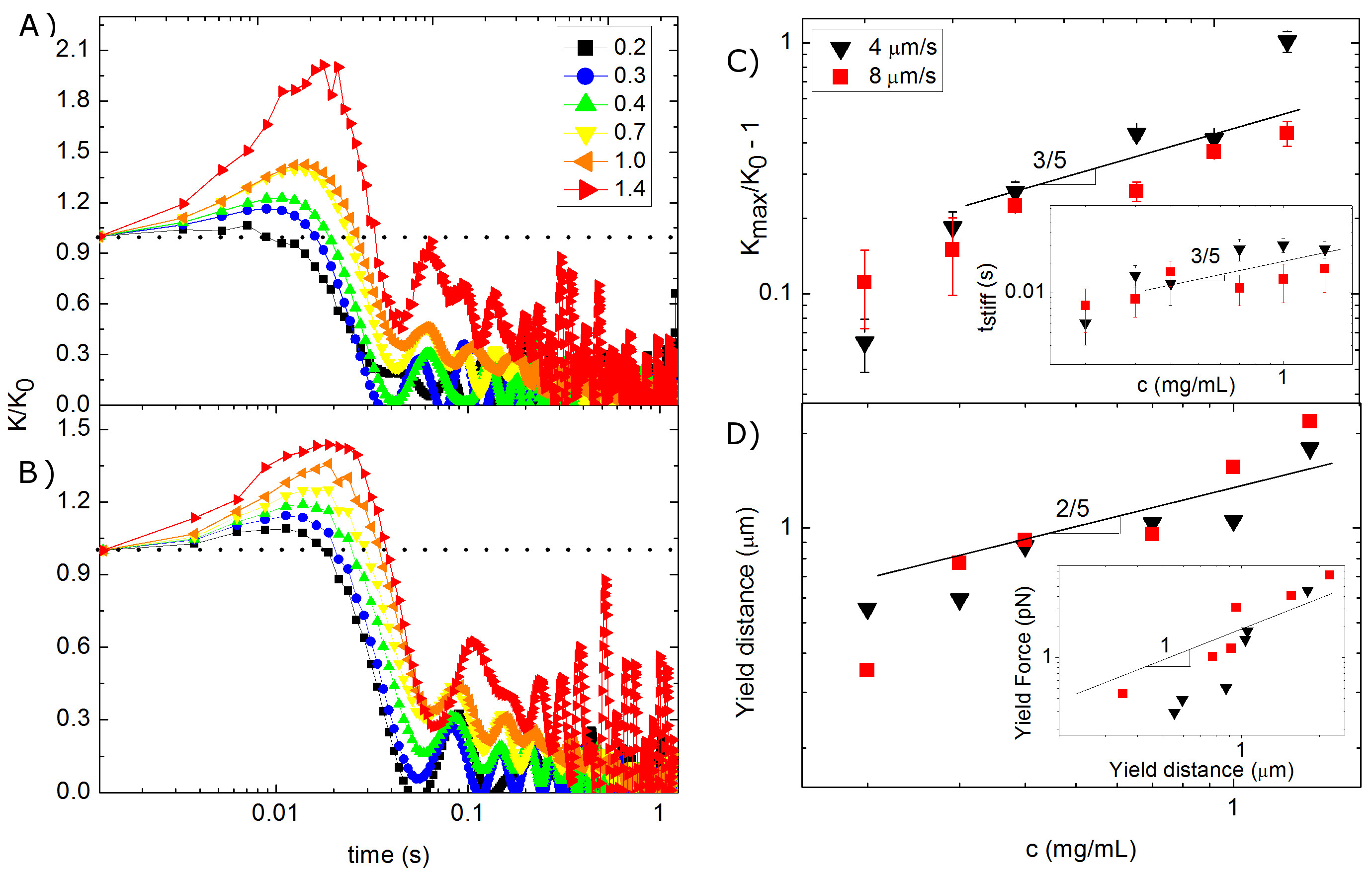}
\caption{ Microscale stress-stiffening and subsequent yielding are controlled by the number of entanglements constraining each filament (A,B) Effective differential modulus of entangled actin networks calculated as the derivative of the resistive force with respect to position ($K = dF/dx$) for 4 ~$\mu$m/s (A) and  8 ~$\mu$m/s ( B) strains. Each modulus is normalized by the corresponding value at time $t$=0 ($K_0$) and color schemes are as in Fig 2A,B. The horizontal dotted line guides the eye to show stress stiffening ($K/K_0 >1$) or softening ($K/K_0 <1$) (C)  Stress-stiffening parameter $(K_{max}/K_0) - 1$, as a function of actin concentration, where $K_{max}$ is the maximum modulus value reached for each concentration. The solid line represents the scaling of the entanglement tube density along the probe path with concentration (i.e. $d_t^{-1}\sim c^{3/5}$). The inset shows the time at which $K_{max}$ is reached ($t_{stiff}$) which also scales as $d_t^{-1}$. Note stiffening only occurs for times comparable to the fastest relaxation timescale of the network ($\tau_{mesh}$). (D)  Yield distance as a function of concentration. Yield distance is determined as the distance at which $K$ drops to $1/2e$ of its initial value $K_0$ and indicates the distance at which the response becomes primarily viscous. The solid line is the predicted scaling for the induced force necessary to disrupt the cohesive entanglement force which scales as $l_{ent}^{-1} \sim c^{2/5}$.The inset further supports the cohesive breakdown model of yielding which predicts $f_y \sim d_y$.}\label{Actin2}
\end{figure}

\begin{figure}[ht!]
\centering
\includegraphics[height=0.8\textwidth]{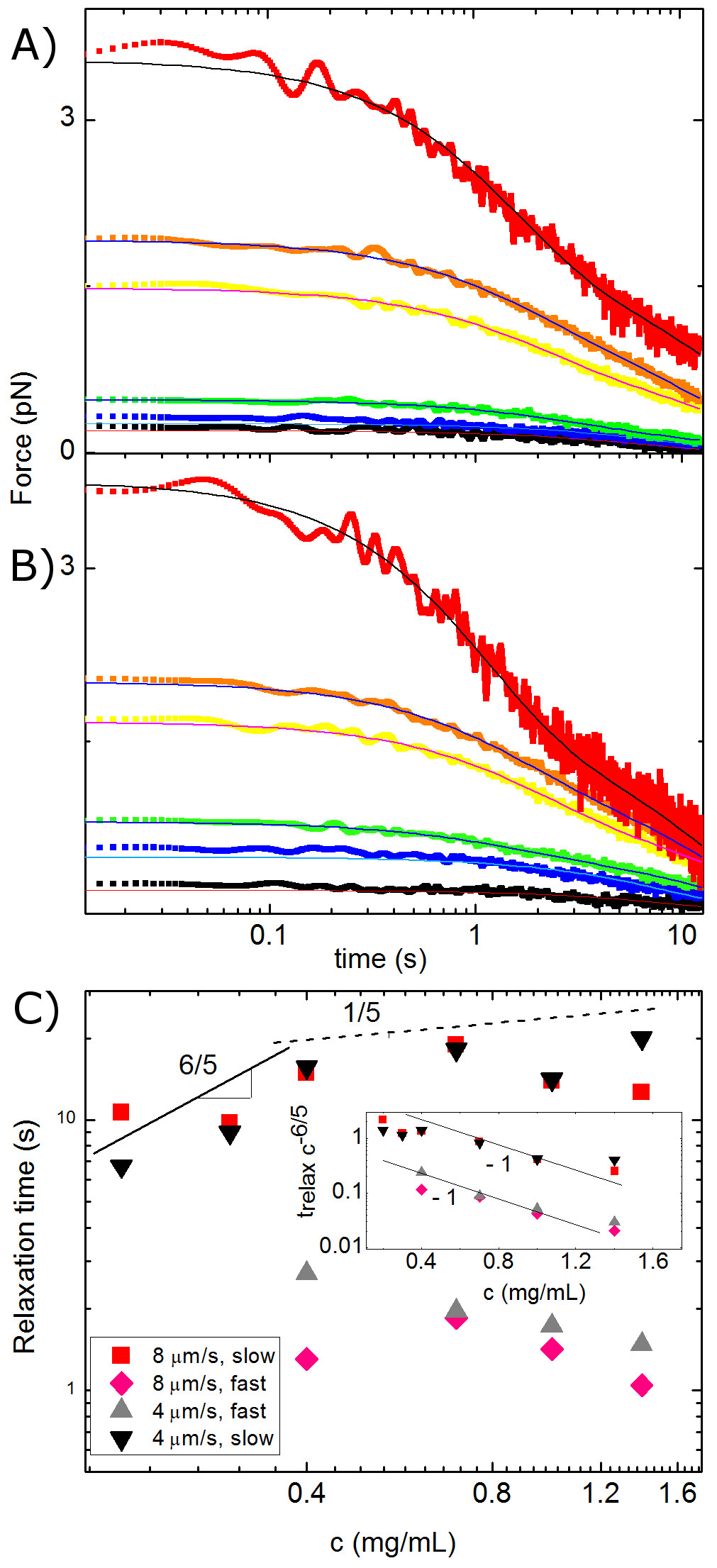}
\caption{Entangled actin force relaxation displays concentration dependent tube dilation and non-classical two-phase relaxation for $c>c^{*}$.(A,B) Relaxation of induced force following the strain as a function of time for (A) 4~$\mu$m/s and (B) 8~$\mu$m/s. The solid lines are fits of the data to exponential decays. The higher four actin concentrations ($c>c^{*}$) can only be described with a double exponential function whereas the lower concentrations (0.2 and 0.3 mg/ml) are well fit with a single exponential function of time. (C) Characteristic force decay times as a function of concentration, determined from the corresponding exponential fits for  both speeds shown in (A) and (B). The solid line indicates the DE predicted scaling of the reptation time with concentration ($\tau_D \sim c^{6/5}$) and the dashed line shows the scaling predicted by the SS model for concentration-dependent tube dilation in the nonlinear regime ($\tau_D \sim c^{1/5}$). The inset shows the measured decay times $t_{relax}$ normalized by the classically predicted reptation time scaling to demonstrate the predicted scaling $\frac{\tau\textprime_D}{\tau_D} \sim c^{-1}$. Higher concentration results demonstrate agreement with SS model predictions for tube dilation and lateral hopping in the nonlinear regime.}\label{Actin3}
\end{figure}

We also measure the relaxation of induced force following the strain (Fig.~\ref{Actin3}(A,B)). For $c < 0.4$ mg/ml force relaxation is well-described by a single exponentially decaying function of time, which implies a single intrinsic relaxation mechanism with a timescale that does not vary with time. As displayed in Fig.~\ref{Actin3}(C), this relaxation time is independent of speed and scaling agrees with the predicted scaling for the reptation time for entangled rigid rod polymers $\tau_{D}\sim c^{6/5}$. This linear response, described by classical DE theory, is in contrast to the displayed relaxations for $c>0.4$ mg/ml networks, which can only be accurately described by a sum of two exponential decays, indicating two distinct relaxation mechanisms with well-separated timescales. The magnitudes of the longer of these two decay times is similar to relaxation times measured for the lower concentrations; however, the scaling with concentration is much weaker than the predicted 6/5 exponent. As described in the Introduction, the SS reptation model extension, which incorporates tube dilation into relaxation dynamics, predicts that tube dilation leads to faster reptation times $\tau\textprime_D$ with the increase from the steady state time scaling as $\frac{\tau\textprime_D}{\tau_D} \sim (\frac{d\textprime_t}{d_t})^{2} \sim c^{-1}$.~\cite{sussman2012microscopic, sussman2013entangled} Combining this prediction with the linear regime rigid rod scaling $\tau_D \sim c^{6/5}$ leads to $\tau\textprime_D \sim c^{1/5}$ which closely describes the concentration scaling of the slower relaxation times with concentration (Fig 4(C) inset). The onset of concentration-dependent tube dilation is coupled with a $\sim$10x faster relaxation mechanism ($\tau_{fast} = \frac{\tau\textprime_D}{10}$) with a similar concentration dependence as the slow timescale. As described in the Introduction, in line with the SS model, we can understand this faster relaxation as due to lateral hopping out of constraining tubes due to fluctuation-induced temporary yielding. The coupled emergence of this lateral hopping mechanism with the concentration-dependent dilation indicates that hopping only plays a significant role in relaxation when entanglement tubes are sufficiently dilated to allow for fluctuation-induced transient yielding of tube constraints.

\section{Conclusion}

Networks of entangled filamentous actin, a key cytoskeleton protein and a model semiflexible polymer, have been shown to display complex viscoelastic mechanics that are highly dependent on the concentration of actin filaments as well as the nature and scale of the applied strain. Here, we have used active microrheology techniques to characterize the concentration dependence of the microscale nonlinear mechanical response of entangled actin. While we previously identified a unique nonlinear response for entangled actin subject to strains faster than $\dot\gamma_c \simeq 3 s^{-1}$, our collective results reported here demonstrate a previously unpredicted and unreported critical concentration (i.e. entanglement density) for nonlinear response features to emerge. Beyond this concentration $c^{*} \sim 0.4$ mg/ml, we show that entangled actin stiffens at timescales comparable to the fastest relaxation timescale of the network $\tau_{mesh}$, and the degree of stiffening $S$ and stiffening timescale scales inversely with the theoretical entanglement tube diameter, i.e. $S\sim d_t^{-1} \sim c^{3/5}$. At longer timescales the network yields to a viscous regime with the distance and corresponding force at which yielding occurs scaling as the number of entanglements along each filament ($f_y \sim d_y \sim l_{ent}^{-1} \sim c^{2/5}$). Stiffening and yielding dynamics are consistent with recent predictions of strain-induced breakdown of the cohesive entanglement force. We further show that above $c*$ force relaxation proceeds via two distinct mechanisms: slow reptation out of dilated tubes coupled with $\sim$10x faster lateral hopping. Tube dilation and commensurate reduced relaxation times scale inversely with concentration, in agreement with recent theoretical predictions. Our results, directly applicable to the larger class of semiflexible polymers, not only resolve important scaling discrepancies reported in the literature but also reveal dynamical scaling laws not previously identified or predicted. Thus, our measurements shed much needed new light on the controversial and scarcely explored microscale dynamics of entangled polymers driven far from equilibrium. 

\begin{acknowledgement}
This research was funded by an NSF CAREER Award, grant number 1255446. 
\end{acknowledgement}

\bibliographystyle{achemso}
\bibliography{Revised Manuscript.bbl}

\end{document}